\documentclass[proof]{pasj00}

\newcommand{\chobi}{\hspace{0.5mm}}

\newcommand{\NH}{{\mbox{\rm\chobi $N_{\rm H}$}}}

\newcommand{\UNITEFLUX}{{\rm ergs~cm$^{-2}$~s$^{-1}$}}

\newcommand{\UNITLUMI}{{\rm ergs~s$^{-1}$}}

\newcommand{\UNITNH}{{\rm cm$^{-2}$}}

\newcommand{\KT}{{\it k$_{\rm B}T$}}

\begin{document}
\SetRunningHead{Tsuru, Nobukawa, Nakajima et al. }{A New SNR Candidate and 
an Associated Outflow in the Sgr~C Region}
\Received{2008/08/05}
\Accepted{2008/09/15}

\title{A New Supernova Remnant Candidate and an Associated Outflow in the Sagittarius~C Region}

\author{%
  Takeshi Go \textsc{Tsuru},\altaffilmark{1}
  Masayoshi \textsc{Nobukawa},\altaffilmark{1}
  Hiroshi \textsc{Nakajima},\altaffilmark{2}
  Hironori \textsc{Matsumoto},\altaffilmark{1}\\
  Katsuji \textsc{Koyama},\altaffilmark{1}
  Shigeo \textsc{Yamauchi},\altaffilmark{3}
}
\altaffiltext{1}{%
   Department of Physics, Graduate School of Science, Kyoto University,
   Sakyo-ku, Kyoto, 606-8502}
\email{tsuru@cr.scphys.kyoto-u.ac.jp}
\altaffiltext{2}{%
  Department of Earth and Space Science, Graduate School of Science, 
  Osaka University, Toyonaka, Osaka, 560-0043}
\altaffiltext{3}{%
  Faculty of Humanities and Social Sciences, Iwate University, 
  3-18-34 Ueda, Morioka, Iwate 020-8550}


\maketitle

\begin{abstract}

We present the Suzaku results on a new candidate of a supernova remnant 
(SNR) in the Sagittarius~C region. 
We detected diffuse X-rays of an elliptical shape (G\,359.41$-$0.12) 
and chimney-like structure (the Chimney), both of which 
are fitted with a thin thermal model of \KT$\sim 1$~keV temperature.
The absorption columns are same between these two structures, 
indicating the both are located at the same distance in the same line of sight.
The narrow band image and one-dimensional profile 
of S\emissiontype{XV} K$\alpha$ at 2.45~keV 
show that the  Chimney is emanating from G\,359.41$-$0.12.  
Therefore these two sources are physically connected with each other. 
The sum of thermal energies of the Chimney and G\,359.41$-$0.12 is 
estimated to be $1.4\times 10^{50}$~ergs, typical for a Galactic SNR. 
G\,359.41$-$0.12 is likely a new SNR candidate 
and the Chimney is an associated outflow.
\end{abstract}

\section{Introduction}

The Galactic center diffuse X-rays (GCDX) is characterized by many
K-shell lines from highly ionized atoms covering a large area of the 
Galactic center (GC) \citep{Koyama89, Yamauchi90, Koyama96, Koyama07c}. 
This indicates the presence of a large scale plasma of temperature 
\KT = 1--10 keV. If this plasma is truly diffuse and extending in the GC, 
it contains a huge thermal energy ($\sim 10^{53-54}$~ergs).
Since the high temperature plasma can not be confined by the Galactic
gravity, it may escape from the GC in a short time ($\sim 10^5$~yr),
the escape time scale.
Thus, we need a huge energy input of $\sim 10^{48-49}$~ergs~yr$^{-1}$.

One plausible hypothesis is a multiple ($\sim 1000$) supernovae 
in the GC region. 
However, only a few supernova remnants (SNRs) had so far been detected 
in the X-ray band. 
This situation has been changing after the launch of the Suzaku satellite. 
The X-ray Imaging Spectrometer (XIS) combined with the X-Ray Telescope 
(XRT) onboard Suzaku has superior performance, a large
collecting area, a stable and low non-X-ray background (NXB) and a high
energy-resolution. 
In order to make the most of these superiority, a significant amount of 
the Suzaku observation time are allocated to the GC. 
As a result, Suzaku has discovered several new SNRs located 
in the Sagittarius (Sgr)~A, Sgr~B1, B2 and Sgr~D 
regions \citep{Koyama07b, Nobukawa08, Sawada08, Mori08a, Mori08b}.

Sgr~C is one of the brightest radio complexes in the GC. 
There are well-defined non-thermal radio filaments (NTFs), 
some compact and evolved H\emissiontype{II} regions, 
far-IR and submillimeter sources, and giant molecular clouds 
\citep{Liszt85, Tsuboi91, Lis94, Liszt95}. 
Accordingly, a rich variety of X-ray features may be expected. 
However, the X-ray studies have been concentrated on the 
diffuse 6.4~keV line emission (Fe\emissiontype{I} K$\alpha$) 
\citep{Murakami01, Yusef07, Nakajima08}; 
few results on thermal diffuse emissions have been reported so far. 
Thus, we conducted the studies of thermal emissions from the Sgr~C 
region with Suzaku. 
\citet{Nakajima08} reported the 6.4~keV clumps in the Sgr C region.
Using the same data set, we searched for thermal features, and
found clumpy emissions of thin thermal nature. We report on the results  
and discuss the origin of the diffuse thermal X-rays.

Following \citet{Nakajima08}, ``the east'' and ``the north'' are referred 
as the positive Galactic longitude side,
and the positive Galactic latitude side, respectively. 
The distance to Sgr~C is assumed to be 8.5~kpc. 
The solar abundance and photoelectric absorption cross-section are
taken from the tables of \citet{Anders89} and \citet{Church92},
respectively.

\section{X-Ray Data Processing}\label{sect:obs}

The Sgr~C region was observed with the XIS from 2006 Feb. 20 to 23, 
and the XIS data were obtained with 
the 3$\times$3 and 5$\times$5 modes and with the normal clocking mode.
The pointing position was R.A.=\timeform{17h44m37s.30}, 
Decl.= \timeform{-29D28'10''.2} (J2000.0).
The XIS consists one (XIS~1) back-illuminated (BI) CCD and three sensors 
(XIS~0, 2 and 3) front-illuminated (FI) CCDs. 
The detailed descriptions of the Suzaku satellite, XRT, and XIS can 
be found in \citet{Mitsuda07}, \citet{Serlemitsos07} and \citet{Koyama07a}, 
respectively.

The data reduction and screenings are essentially the same as 
\citet{Nakajima08}, but we applied the calibration database 
(CALDB) released on 
2008 July 2\footnote{http://www.astro.isas.jaxa.jp/suzaku/caldb/}. 
For the data processing, we selected X-ray events by subtracting the 
NXB from the raw data. 
Since the flux of the NXB depends on the geomagnetic cut-off rigidity (COR) 
\citep{Tawa08}, 
we applied the COR-sorted NXB (using the data sets provided by the XIS 
team\footnote{http://www.astro.isas.jaxa.jp/suzaku/analysis/xis/nte/}), 
to become the same COR distribution as that in the Sgr~C observation. 
We exclude the data of the corners in the field of view (FOV) 
illuminated by the onboard calibration sources of $^{55}$Fe.

\section{Analysis and Results}

\subsection{Imaging of Thermal Emissions}\label{sec:imaging ana}

\begin{figure}
  \begin{center}
    \FigureFile(80mm,80mm){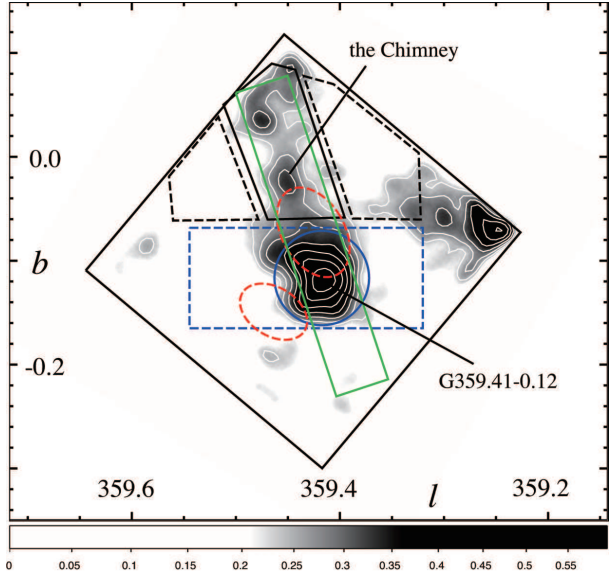} 
  \end{center}
  \caption{
    The 2.45~keV-line (S\emissiontype{XV} K$\alpha$) image 
    in the energy band of 2.40--2.50~keV after the smoothing 
    with the Gaussian kernel of 60~pixels ($\timeform{1'.04}$). 
    Contour levels are set at 
    $(4.2$, $4.9$, $5.7$, $6.6$, $7.8$, $9.0$, 
    $10.6$, $12.3$, $14.4$, $16.8) 
    \times 10^{-7}$ ph~s$^{-1}$~cm$^{-2}$~arcmin$^{-2}$. 
    The typical 1$\sigma$ error is 
    0.5$\times 10^{-7}$ ph~s$^{-1}$~cm$^{-2}$~arcmin$^{-2}$. 
    We collected the source and background spectra of the Chimney 
    from the regions shown with solid blue and dashed black lines, 
    respectively. 
    Extraction of the source spectra of G\,359.41$-$0.12 
    were made from the region given with the solid line, 
    while its background spectra were obtained from the region 
    of dashed blue lines but the source area (solid bule circle) 
    was excluded. 
    Also the data within the red dashed ellipses (M\,359.43$-$0.07 and 
    M\,359.47$-$0.15) were excluded to minimize the contamination 
    from the strong 6.4~keV emission line and its associated continuum. 
    The green box with the size of $\timeform{18'.5}\times \timeform{3'}$ 
    gives the region for the one-dimensional profile shown in 
    figure~\ref{fig:profile}.
    }
  \label{fig:2.45keV_image}
\end{figure}

We searched for a thin thermal plasma in the images of
S\emissiontype{XV} K$\alpha$ lines at 2.45 keV
(for a plasma temperature of \KT $\lesssim 1$~keV)
and/or Fe\emissiontype{XXV} K$\alpha$ lines at 6.68 keV
(for a plasma temperature of \KT $\gtrsim 1$~keV).
For the data processing and analysis of the imaging study, we co-added
the four CCD (XIS~0, 1, 2, and 3) data and made the exposure and
vignetting corrections.
In the Fe\emissiontype{XXV} K$\alpha$ image (6.6--6.8~keV), 
we found no significant structure.
However, in the narrow band image of 
the S\emissiontype{XV} K$\alpha$ line at 2.45 keV (2.40--2.50~keV), 
we found a bright extended source at 
$(l,b)=(\timeform{359D.41}, \timeform{-0D.12})$, 
hence designated as G\,359.41$-$0.12 hereafter 
(figure~\ref{fig:2.45keV_image}). 
Another diffuse emission has a chimney-like structure extending 
from G\,359.41$-$0.12 to the north direction, 
which we call as ``the Chimney''. 
Two known bright point sources, 1E\,1740.7$-$2942 
($\sim 5\times 10^{-10}$\UNITEFLUX) 
and A1742-294 ($\sim 3$--$6\times 10^{-10}$\UNITEFLUX), are 
located outside the FOV in the western and south-east directions, 
respectively (e.g., Sidoli et al. 1999). 
The contamination due to the stray-light from the two sources 
is only $\sim 2-3$\% of the flux of the Chimney in the energy band of 
2.40--2.50~keV \citep{Serlemitsos07}. 
Thus, the Chimney is a real structure. 

\begin{figure}
  \begin{center}
    \FigureFile(80mm,80mm){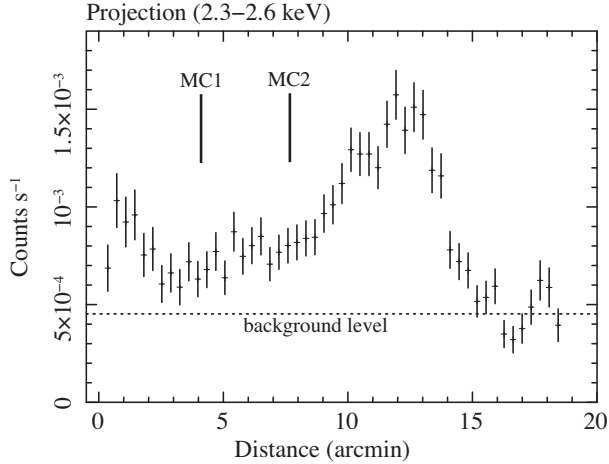} 
  \end{center}
  \caption{
    One-dimensional surface brightness profile at 2.45~keV (2.30--2.60~keV) 
    along the green box connecting the Chimney with G\,359$-$0.12 
    (figure~\ref{fig:2.45keV_image}). 
    The origin of the profile is defined at the northern side of 
    the rectangle. 
    The dotted line shows the background level obtained from the 
    the background regions for the Chimney and G\,359.41$-$0.12
    indicated in figure~\ref{fig:2.45keV_image}. 
    ``MC\,1'' and ``MC\,2'' show the positions of ridges of the 
    molecular clouds crossing the Chimney 
    (see figure~\ref{fig:13CO_2.45keV_image}). 
    }
  \label{fig:profile}
\end{figure}

To see if the Chimney is connected with G\,359.41$-$0.12 quantitatively,
we constructed an one-dimensional surface brightness profile 
at 2.45~keV with the length of $\timeform{18'.5}$ 
along the green rectangle in figure~\ref{fig:2.45keV_image}. 
The origin of the profile is defined at the northern side of the box. 
The background level was obtained from the background regions of 
the spectra of the Chimney and G\,359.41$-$0.12 shown in 
figure~\ref{fig:2.45keV_image}. 
The 2.45~keV profile shown in figure~\ref{fig:profile} clarifies that 
the surface brightness from G\,359.41$-$0.12 to the Chimney are well 
above the background level. 
On the other hand, the profile is not smooth but has two dips at 
distances of $\sim \timeform{3'.5}$ and $\sim \timeform{7'}$. 

As mentioned above, a bright source 1E\,1740.7$-$2942 
is located outside the FOV in the western direction. 
A possible radio SNR having a size of 22~pc$\times$10~pc, 
G\,359.07$-$0.02, is cataloged in \citet{LaRosa00}. 
Thus, the western excess would be due to the XRT PSF (point spread function) 
tail of 1E\,1740.7$-$2942 and/or eastern edge of G\,359.07$-$0.02. 
At the time of writing this paper, 
Suzaku already made a pointing observation of the western adjacent region. 
However, the exposure time was only $\sim 20$~ksec, 
and hence it is not clear whether the excess near 
the western edge of the FOV is a real excess or a spill over of 
these X-ray sources. We therefore will not discuss on this excess in 
this paper.

\subsection{Spectral Analyses}

In the spectral studies, we treated the co-added FIs (XIS~0,2,3) data
and BI (XIS~1) data separately because the response function of FIs and
BI are significantly different.
The spectra were extracted from the regions 
shown in figure~\ref{fig:2.45keV_image}.
Since the surface brightness of the GCDX, the most significant
background component, depends on the Galactic latitude, the background
regions for the Chimney and G\,359.41$-$0.12 were selected to have the
same latitude as those of corresponding source areas.
The energy-dependent vignetting were corrected by multiplying the 
effective-area ratio of the source region to the corresponding 
background area for each energy bin. 
Since the 6.4 keV clumps, M\,359.43$-$0.07 and M\,359.47$-$0.15 have 
a strong 6.4~keV-line and associated continuum emission \citep{Nakajima08},
we excluded the regions of these 6.4~keV clumps from both of the source and
background areas, in order to minimize the contamination of the clumps.

\begin{figure}
  \begin{tabular}{cc}
    \begin{minipage}{0.5\textwidth}
      \FigureFile(80mm,80mm){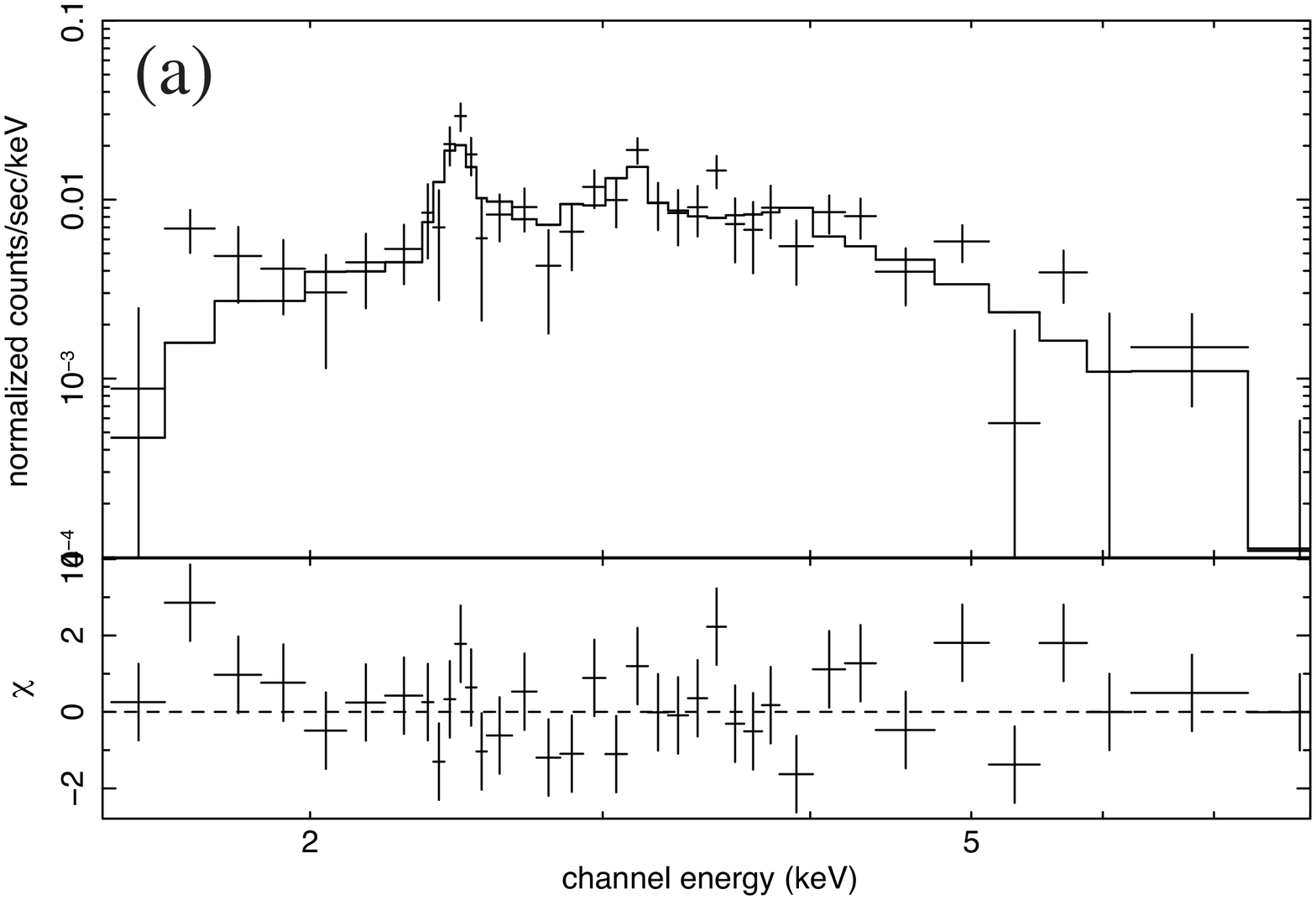} 
    \end{minipage}
    \begin{minipage}{0.5\textwidth}
      \FigureFile(80mm,80mm){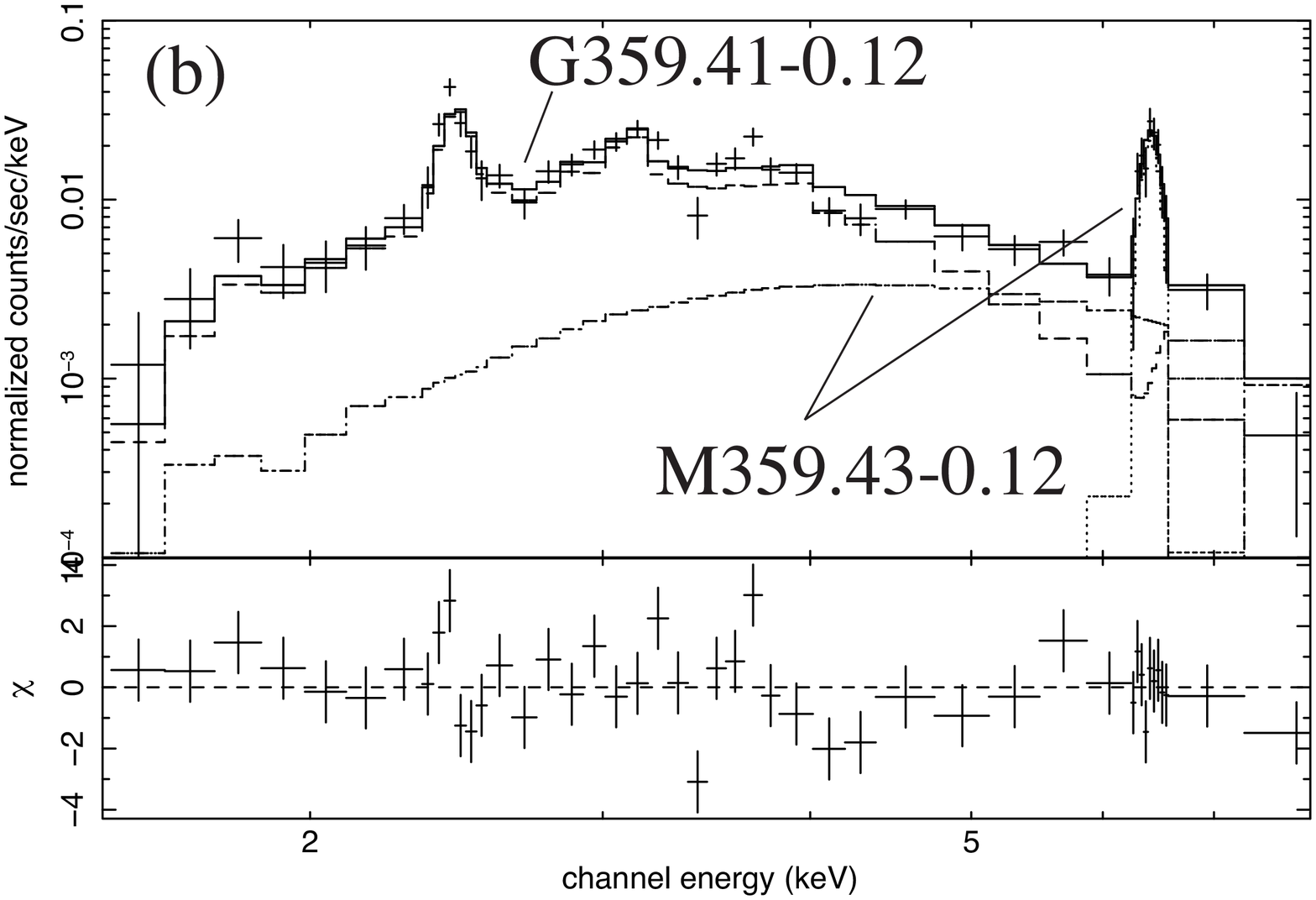} 
    \end{minipage}
  \end{tabular}
  \caption{The background subtracted FIs spectra of (a) the Chimney 
    and (b) G\,359.41$-$0.12 with the best fit model spectra. 
    Although, we simultaneously fitted the FIs and BI spectra, 
    only the FI results are given for brevity.}
  \label{fig:spec}
\end{figure}

The background (the GCDX) subtracted FIs spectrum of the Chimney is shown 
in figure~\ref{fig:spec}~(a). 
We see two conspicuous emission 
lines from highly ionized atoms, S\emissiontype{XV} K$\alpha$ (2.45~keV) 
and Ar\emissiontype{XVII} K$\alpha$ (3.12~keV). 
We, therefore, fitted the spectrum by a thin thermal plasma model in
collisional ionization equilibrium affected by photoelectric
absorption (\verb/vapec/ and \verb/wabs/ models in XSPEC, respectively). 
The fit was reasonably good as is shown in table~\ref{tab:fitresults2}. 

The background (the GCDX) subtracted spectra of G\,359.41$-$0.12 shows 
prominent 6.4~keV emission lines from Fe\emissiontype{I} K$\alpha$ as well as
S\emissiontype{XV} K$\alpha$ and Ar\emissiontype{XVII} K$\alpha$. 
This 6.4~keV line emission come from a 6.4~keV clump, M\,359.43$-$0.12, 
which is located inside the source region of G\,359.41$-$0.12 and 
hence difficult to specially exclude \citep{Nakajima08}. 

The bight 6.4~keV clumps of M\,359.43$-$0.07 and M\,359.47$-$0.15,
locating near G\,359.41$-$0.12 and M\,359.43$-$0.12, 
have essentially the same spectrum \citep{Nakajima08}.  
Thus, assuming M\,359.43$-$0.12 also has the same spectrum, 
we added a model of the fixed absorption column, photon
index and equivalent widths of the best-fit of  
M\,359.43$-$0.07 and M\,359.47$-$0.15, and fit with the same model 
as the Chimney (a thin thermal model). 
The fit was acceptable as is shown in table~\ref{tab:fitresults2}.
We further examined how the results of G\,359.41$-$0.12 
depend on the spectral shape of the M\,359.43$-$0.12, by 
changing the model parameters within the errors.
Then we found no significant change of the results; 
only the errors of the best-fit parameters of G\,359.41$-$0.12 become larger  
by a factor of $\sim 1.5$.


\begin{table}
  \begin{center}
    \caption{Best-fit parameters of the Chimney and G\,359.41$-$0.12}
    \label{tab:fitresults2}
    \begin{tabular}{lcc}
      \hline
      Parameters 
          & the Chimney 
          & G\,359.41$-$0.12 \\
      \hline
      \ \ $N_{\rm H}$ ($10^{22}$~cm$^{-2}$) 
          & 10.1 (8.6--12.2) 
          & 12.0 (11.5--12.4) \\
      \ \ \KT\ (keV) 
          & 1.15 (0.93--1.37) 
          & 0.90 (0.84--0.94) \\
      \ \ $Z_{\rm S}$, $Z_{Ar}$ ($Z_\odot$) \footnotemark[$\dagger$] 
          & 1.73 (1.11--2.65)  
          & 1.72 (1.51--2.07) \\
      \hline
      $f_{\rm X}$\footnotemark[$\sharp$] (thermal) (\UNITEFLUX )
          & $3.23\times 10^{-13}$ 
          & $3.11\times 10^{-13}$ \\
      $f_{\rm X}$\footnotemark[$\sharp$] (power-law) (\UNITEFLUX )
          & $\cdots$
          & $2.39\times 10^{-13}$ \\
      \hline
      $\chi ^{2}$/d.o.f.  
          & $74.9/55$ 
          & $128.4/91$  \\
      \hline
      \multicolumn{3}{@{}l@{}}{\hbox to 0pt{\parbox{100mm}{\footnotesize
        Notes. -- 
          The values in parentheses represent the 90\% confidence intervals. 
            \par\noindent
        \footnotemark[$\dagger$]The values of $Z_{\rm S}$ and 
           $Z_{\rm Ar}$ are fixed to one another. 
           $Z$ of the other elements are fixed to the solar values. 
         \footnotemark[$\sharp$]Observed (absorption not corrected) 
            flux at the 1.5--8.0~keV band.
            \par\noindent
          }\hss}}
    \end{tabular}
  \end{center}
\end{table}

\section{Discussion}

We discovered two diffuse X-ray 
sources with the 2.45~keV emission line 
(S\emissiontype{XV} K$\alpha$), G\,359.41$-$0.12 and the Chimney. 
There is a well-defined non-thermal radio filament, G\,359.45$-$0.06, 
running north-western direction in the Sgr~C region 
\citep{Liszt85, Liszt95}.
However, the position angle of G\,359.45$-$0.06 disagree with 
that of the Chimney by $\sim \timeform{40D}$, and hence the Chimney is 
not related to G\,359.45$-$0.06 (figure~\ref{fig:13CO_2.45keV_image}). 
In the followings, we discuss the structure and origin of the plasmas. 

\subsection{Plasma Parameters}

We first derived physical parameters of the Chimney and G\,359.41$-$0.12, 
summarized in table~\ref{tab:physparam}.
We assumed the Chimney and G\,359.41$-$0.12 have the plasmas of an 
cylinder with the diameter of $7.4$~pc (\timeform{3.0'})
and length of $20$~pc (\timeform{8.0'}), and 
an ellipsoid with the dimensions of 
8.6~pc$\times$ 8.6~pc $\times$ 12.4~pc 
($\timeform{3.5'}\times\timeform{3.5'}\times \timeform{5.0'}$), 
respectively. 

The plasma temperatures give sound velocities of 
$C_{\rm s}=560$~km~s$^{-1}$ and $480$~km~s$^{-1}$ for 
the Chimney and G\,359.41$-$0.12, respectively. 
Hence, dividing the length or major axis by the sound 
velocities, we can estimate the dynamical time scales of 
the Chimney and G\,359.41$-$0.12 to be $4.0\times 10^4$~yr 
and $2.5\times 10^4$~yr. 

\begin{table}
  \begin{center}
    \caption{Physical parameters of the Chimney and G\,359.41$-$0.12. 
      \footnotemark[$*$]}
    \label{tab:physparam}
    \begin{tabular}{lcc}
      \hline
      Parameters 
          & the Chimney             
          & G\,359.41$-$0.12 \\     
      \hline
      $\int n_{e}n_{\rm H} dV$~(cm$^{-3})$
          & $4.9\times 10^{57}$ 
          & $1.3\times 10^{58}$ \\
      $V$ (pc$^{3}$) 
          & $8.5\times 10^2$
          & $4.8\times 10^2$ \\
      $n_{\rm e}$ (cm$^{-3}$) 
          & $0.49f^{-1/2}$
          & $1.1f^{-1/2}$ \\
      $M$ (\MO) 
          & $14f^{1/2}$
          & $14f^{1/2}$ \\
      $E_{\rm th}$ (ergs) 
          & $7.6\times 10^{49}f^{1/2}$
          & $5.9\times 10^{49}f^{1/2}$ \\
      \hline
      $L_{\rm X}$\footnotemark[$\natural$] (\UNITLUMI ) 
          & $2.2\times 10^{34}$
          & $3.9\times 10^{34}$ \\
      \hline
      \multicolumn{3}{@{}l@{}}{\hbox to 0pt{\parbox{80mm}{\footnotesize
        Notes. -- The distance is assumed to be 8.5~kpc. 
        $M$, $E_{\rm th}$ and $f$ are the mass, thermal energy 
        and filling factor of the plasma, respectively. 
        The other notations refer to table~\ref{tab:fitresults2}. 
        \footnotemark[$\natural$]Unabsorbed luminosity between 1.5-8.0~keV. 
          }\hss}}
    \end{tabular}
  \end{center}
\end{table}

\subsection{Physical Connection of the Chimney and G\,359.41$-$0.12}
\label{sec:phys connect}

We discuss if the plasmas of the Chimney and G\,359.41$-$0.12 
are connected with each other. 
The one-dimensional profile in figure~\ref{fig:profile} suggests that the 
X-ray emission continuously distributes between the two objects without
break.
The absorption columns of the Chimney and G\,359.41$-$0.12 
are almost identical at \NH=(10-12)$\times 10^{22}$~cm$^{-2}$,
suggesting that these two sources are at the same distance in the 
same line of sight.
The temperatures and metal abundances of these sources 
are essentially the same. 
Therefore, we can infer that the Chimney and G\,359.41$-$0.12 are
physically connected with a similar nature.

\subsection{Absorption by Molecular Clouds}

\begin{figure}
  \begin{center}
    \FigureFile(80mm,80mm){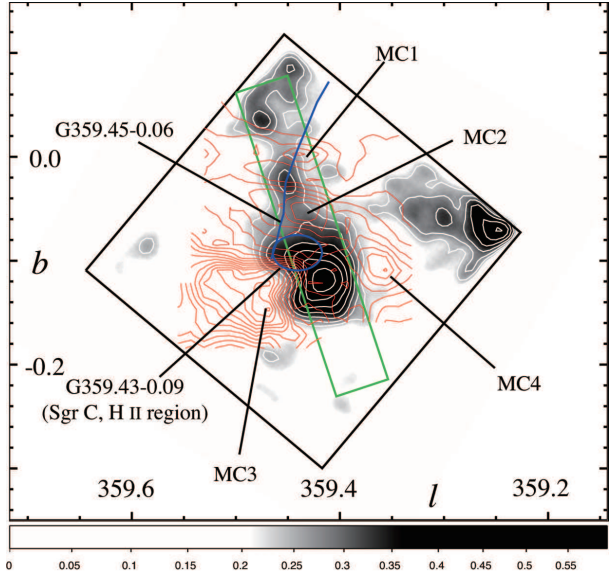} 
  \end{center}
  \caption{
    The contours of $\atom{CO}{}{13}$ at $-84$ to $-24$~km~s$^{-1}$ (red)
    and the schematic diagram of the radio continuum sources of 
    G\,359.45$-$0.06 and G\,359.43$-$0.09 (blue) are 
    overlaid on the 2.45~keV-line (S\emissiontype{XV} K$\alpha$) image. 
    The radio data are adopted from \citet{Liszt95}. 
    MC\,1, 2, 3, and 4 indicate the molecular clouds. 
    The region for the one-dimensional profile in 
    figure~\ref{fig:profile} is shown with the green box 
    ($\timeform{18'.5}\times \timeform{3'}$). 
    }
  \label{fig:13CO_2.45keV_image}
\end{figure}

Figure~\ref{fig:13CO_2.45keV_image} shows $\atom{CO}{}{13}$ emission over the 
range $-84$ to $-24$~km~s$^{-1}$ overlaid on the X-ray image at 2.45~keV 
\citep{Liszt95}. 
Anti-correlation among the four molecular clouds (MC\,1, 2, 3, 4) and 
X-ray emission is clearly seen. 
The positions of MC\,1 and MC\,2 agree well with the two dips seen 
in the one-dimensional profile (figure~\ref{fig:profile}). 
The hydrogen column densities of MC\,1 and MC\,2 estimated from the
$\atom{CO}{}{13}$ emission are \NH$\sim 2\times 10^{22}$~\UNITNH\ 
($\sim 40$\% absorption at 2.45~keV). 
The unabsorbed counting rates plus the background level are obtained  
to be $\sim 8\times 10^{-4}$ and $\sim 1.0\times 10^{-3}$~counts~s$^{-1}$ 
at the position of MC\,1 and MC\,2, respectively. 
Thus, we can assume that MC\,1 and MC\,2 are located in front of 
the Chimney and absorb the X-ray emission; the dips would be due to 
the absorption by the molecular clouds and 
the unabsobed one-dimensional profile in the Chimney would 
decrease monotonically from G\,359.41$-$0.12 without a significant dip.
This result enforces the idea that G\,359.41$-$0.12 and the Chimney 
are physically connected with each other (section~\ref{sec:phys connect}).

Figure~\ref{fig:13CO_2.45keV_image} also shows that the surface brightness
at the 2.45~keV band is low at the positions of MC\,3 and MC\,4. 
The column densities of MC\,3 and MC\,4 obtained from the $\atom{CO}{}{13}$
emission are \NH$\sim 7\times 10^{22}$~cm$^{-2}$ and 
$\sim 3\times 10^{22}$~cm$^{-2}$, respectively. 
Thus, one can explain the anti-correlation 
would be due to the absorption by MC\,3 ($\sim 80$\%) 
and MC\,4 ($\sim 50$\%), as applied to MC\,1 and MC\,2. 
On the other hand, 
since the shapes of MC\,3 and MC\,4 trace the rim of G\,359.41$-$0.12 well, 
the other possibility is that the molecular clouds block the plasma to 
expand further out. 
Since the surface brightness in the region of MC\,3 and MC\,4 is very low, 
further deep observation is necessary to clarify these two possibilities.

\subsection{Origin of the Plasmas}

First, we discuss the origin as star clusters. 
The observed X-ray luminosities of G\,359.41$-$0.12 and the Chimney are
(2--4)$\times 10^{34}$~\UNITLUMI\ in the 1.5--8~keV band. These luminosities 
are comparable with or higher than those of the Arches 
($\sim 3\times 10^{34}$~\UNITLUMI) and Quituplet 
($\sim 4\times 10^{33}$~\UNITLUMI) clusters, 
the X-ray brightest star clusters in the Galaxy 
\citep{Tsujimoto07, Wang06}. 
The canonical metal abundance of the Galactic H\emissiontype{II} 
regions obtained with X-ray spectroscopies is 0.3 solar 
(e.g., Getman et al. 2005), 
while G\,359.41$-$0.12 and the Chimney have metal abundances of 
$1\sim 2$ solar. 
There is a radio bright H\emissiontype{II} region, G\,359.43$-$0.09
(Sgr~C) as shown in figure~\ref{fig:13CO_2.45keV_image}
\citep{Liszt95}. However, the peak position of G\,359.41$-$0.12 is 
$\timeform{2'.4}$ shifted from that of G\,359.43$-$0.09. 
Thus, it is unlikely that the X-rays of G\,359.41$-$0.12 and the Chimney
come from  the H\emissiontype{II} regions.

A pulsar wind nebula and jet from a compact source are also unlikely 
as the origin of the Chimney and G\,359.41$-$0.12, because a power law 
model failed to fit the observed spectra. 

The X-ray luminosity, temperature, and the other physical parameters of
G\,359.41$-$0.12 and the Chimney  are consistent with those of the SNRs 
detected in the GC \citep{Nobukawa08, Mori08a, Sawada08, Mori08b}. 
The sum of thermal energies of the Chimney and G\,359.41$-$0.12 is
$1.4\times 10^{50}$~ergs, typical for a Galactic SNR. 
Including the possible energy of the plasma bulk motion 
with the sound velocity into the total energy, 
it is still consistent with a single supernova. 
Therefore, we propose G\,359.41$-$0.12 is a new SNR candidate 
and the Chimney is an outflow plasma from the SNR. 
The molecular clouds of MC\,3 and MC\,4 can block the plasma from
expanding except in the northern direction. 
Then, the outflow in the direction of the north may be formed. 
In any case, the morphology of single SNR plus a chimney-like thermal 
outflow with the dimension of $\sim$ 8~pc $\times$ 8~pc $\times$ 34~pc is 
very unusual as a SNR.
Thus, further deep observation is highly required. 

\bigskip
\bigskip

We are grateful to all members of the Suzaku hardware and software
teams and the science working group.
This work was supported by the Grant-in-Aid for the Global COE Program
``The Next Generation of Physics, Spun from Universality and
Emergence'' from the Ministry of Education, Culture, Sports, Science
and Technology (MEXT) of Japan.
This work was also supported by Grants-in-Aid of Ministry of
Education, Culture, Sports, Science and Technology No. 18204015 and 
20340043. 
MN and HN are financially supported by the Japan Society for the
Promotion of Science.

\end{document}